\documentstyle[aps,psfig,floats,epsf,twocolumn,prl]{revtex}

\begin{document}
\tighten
\draft
\twocolumn[
\hsize\textwidth\columnwidth\hsize\csname @twocolumnfalse\endcsname  

\title{Nonresonant $B_{\rm 1g}$ Raman scattering in the Hubbard model}
\author{J.K. Freericks$^*$, T.P. Devereaux$^\dagger$ and R. Bulla$^\ddagger$}
\address{$^*$Department of Physics, Georgetown University, Washington, DC
20057, U.S.A.}
\address{$^\dagger$Department of Physics, University of Waterloo, Canada}
\address{$^\ddagger$Theoretische Physik III, Elektronische Korrelationen und 
Magnetismus, Institut f\"ur Physik, Universit\"at Augsburg,  D-86135 Augsburg,
Germany}
\date{\today}
\maketitle

\widetext
\begin{abstract}
The numerically exact solution for nonresonant $B_{\rm 1g}$
Raman scattering is presented for the half-filled
Hubbard model in infinite dimensions.
This solution illustrates the modifications of the Raman response (in
a system tuned through the quantum-critical point of a metal-insulator
transition) due to Fermi-liquid properties in the metallic phase.  In the
insulating phase, we recover the predicted universal behavior, while we find
the Raman response is quite anomalous on the metallic side of the transition.
Our calculated results are similar to those measured in FeSi, SmB$_6$, and
underdoped cuprates.
\end{abstract}
]
\narrowtext

Raman scattering
has been applied to study the excitations of electrons in metals, insulators,
semiconductors, and superconductors. 
Inelastic light scattering reveals symmetry selective properties of the
electron dynamics over a wide range of energy scales and temperatures
(the symmetry is selected by polarizing the incident and reflected light).  
Three principle channels
are usually examined: (i) $A_{\rm 1g}$ which has the full symmetry of
the lattice (i.e. is $s$-like); (ii) $B_{\rm 1g}$ which is a $d$-like symmetry
(that probes the Brillouin-zone axes); and (iii) $B_{\rm 2g}$ which is another
$d$-like symmetry (that probes the Brillouin-zone diagonals).
Shastry and Shraiman (SS) proposed a simple relation~\cite{ss} that 
connects the nonresonant
Raman response to the optical conductivity, which was recently
proven in infinite-dimensions~\cite{freericks_devereaux_long} for 
the $B_{\rm 1g}$ channel.  We employ
the SS relation here to evaluate Raman scattering in the Hubbard model.

\begin{figure}
\epsfxsize=3.0in
\centerline{\epsffile{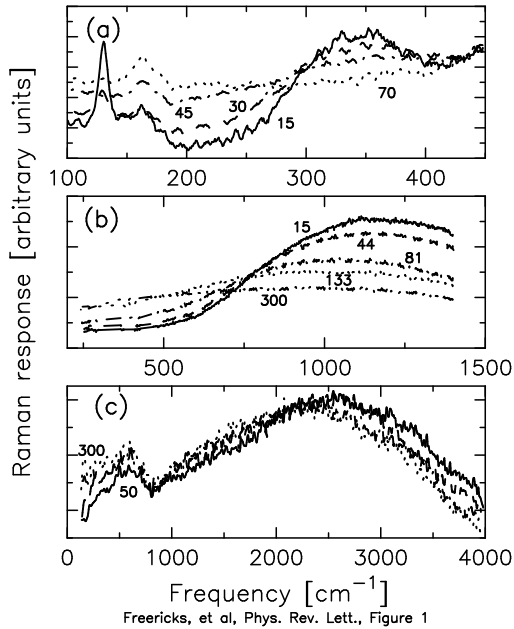}}
\caption{
Experimental Raman response for correlated materials (a) 
SmB$_6$~[3]; (b) FeSi~[4]; and (c) 
underdoped La$_{2-x}$Sr$_x$CuO$_4$~[5] with $x=0.08$.
All of the experimental data show the development of a low-temperature
isosbestic point, which occurs due to the transfer of spectral weight
from low energy to high energy as the temperature is lowered, indicating
the proximity to the quantum-critical point of a metal-insulator transition.
The individual curves are labeled by the temperature in K where the measurement
was taken. In panel (c) only the high temperature (300~K) and the low 
temperature (50~K) are labeled.  The two intermediate curves are at
100 and 200~K respectively.
\label{fig: experiment} }
\end{figure}

Strongly correlated systems as disparate 
as mixed-valence compounds~\cite{SLC1} (such as SmB$_{6}$), 
Kondo-insulators~\cite{SLC2}
(such as FeSi), and the underdoped cuprate high temperature
superconductors~\cite{irwin,hackl,uiuc}, 
show temperature-dependent $B_{\rm 1g}$
Raman spectra that are both remarkably similar and quite anomalous
(see Figure~\ref{fig: experiment}).
This ``universality''
suggests that there is a common mechanism governing the electronic
transport in correlated insulators. As these materials are cooled, they all 
show an increase in the spectral weight in a ``charge-transfer'' peak
with a simultaneous reduction of low-frequency spectral weight.
This spectral weight transfer is slow at high temperatures and then
rapidly increases as the temperature is lowered in the proximity of a
quantum-critical point (corresponding to a metal-insulator transition). 
The Raman spectral range is also
separated into two regions: one where the response decreases as $T$
is lowered and one where the response increases.  The characteristic
frequency that divides these two regions is called the isosbestic point, which 
is the frequency where the Raman response is independent of
temperature.  These anomalous features are
not typically seen in either the $A_{\rm 1g}$ or the $B_{\rm 2g}$ channels.

This anomalous experimental behavior is shown in Figure~\ref{fig: experiment}.
The top panel shows SmB$_6$, which has the added feature of developing
a sharp peak at 130~cm$^{-1}$ (that does not disperse in frequency) when
the temperature is lower than 30~K.  The FeSi data is shown in the middle
panel.  It displays the cleanest signature of these anomalous features. Note
how the isosbestic point only develops at temperatures below 150~K.  The bottom
panel shows the data in the LSCO high-temperature superconductor (which we
have smoothed using a 4253H,twice smoother).  The
isosbestic point is somewhat harder to see here (because of the noise in the
data), but it does clearly develop at about 2000~cm$^{-1}$
as the temperature is lowered (one cannot go below the 
superconducting transition temperature, since additional signatures of
the ordered phase would be seen in the Raman response).  

Theory has just begun to catch up with experiment for electronic Raman 
scattering in strongly correlated materials that are tuned to lie
near a metal-insulator transition~\cite{raman_us,freericks_devereaux_long}.
This theoretical treatment involved
applying the dynamical mean-field theory to the simplest many-body system
that has a metal-insulator transition---the spinless Falicov-Kimball 
model~\cite{falicov_kimball}. While results for the $B_{\rm 1g}$ channel were
shown to be universal on the insulating side of the transition (and
qualitatively agree with the anomalies seen in experiment), on the
metallic side, the effects of a low-temperature Fermi coherence peak 
are unknown, since the Falicov-Kimball model is a non Fermi liquid. 
The iterated-perturbation-theory
approximation has been applied to the Hubbard model before~\cite{craco}
but not in the regime of the metal-insulator transition;
here we provide the numerically exact solution in this regime.

The nonresonant Raman scattering in the $B_{\rm 2g}$ channel 
vanishes~\cite{freericks_devereaux_long}, and in the $A_{\rm 1g}$ channel
it requires knowledge of the local irreducible charge vertex (which is
problematic to calculate in the Hubbard model due to its large
dynamic range, which cannot be treated by maximum-entropy
techniques), so we provide results only for the $B_{\rm 1g}$ sector here.
The Hubbard Hamiltonian
contains two terms:
the electrons can hop between nearest neighbors
[with hopping integral $t^*/(2\sqrt{d})$ on a $d$-dimensional hypercubic
lattice~\cite{metzner_vollhardt}],
and they interact via a screened Coulomb interaction $U$ when they sit on the
same site.  All energies are measured in units of $t^*$.  The Hamiltonian is
\begin{equation}
H =-\frac{t^*}{2\sqrt{d}}\sum_{\langle i,j\rangle, \sigma}c_{i\sigma}^{\dagger}
c_{j\sigma} +
U\sum_in_{i\uparrow}n_{i\downarrow},
\label{eq: ham}
\end{equation}
where $c_{i\sigma}^{\dagger}$ $(c_{i\sigma})$ is the creation
(annihilation) operator for an electron at lattice site $i$ with spin
$\sigma$ and $n_{i\sigma}=c_{i\sigma}^{\dagger}c_{i\sigma}$ is the
electron number operator.  We adjust a chemical potential
$\mu$ to fix the average filling of the electrons. 

The nonresonant Raman response in the B$_{\rm 1g}$ channel has
no vertex corrections~\cite{khurana,raman_us,freericks_devereaux_long} 
and is equal to the bare bubble.  The formula for the Raman response
has been presented
elsewhere~\cite{freericks_devereaux_long} and can also be found from the
SS relation~\cite{ss,jarrell_oc}---the imaginary part of the nonresonant 
$B_{\rm 1g}$ Raman response is  
\begin{eqnarray}
{\rm Im}R(\nu)&=&c\int d\omega [f(\omega)-f(\omega+\nu)]\cr
&\times&\int d\epsilon \rho(\epsilon) A(\epsilon,\omega)A(\epsilon,\omega+\nu),
\label{eq: raman}
\end{eqnarray}
where $f(\omega)=1/[1+\exp(\omega/T)]$ is the Fermi function, 
$A(\epsilon,\omega)={\rm Im}[-1/\pi\{\omega+\mu-\Sigma(\omega)-\epsilon\} ]$
is the spectral function, $\rho(\epsilon)$ is the noninteracting density
of states (a Gaussian here), and $c$ is a constant.  

We study the evolution of the Raman response at
half filling, since one can tune the system to move right through the
quantum-critical point of the metal-insulator transition.  We analyze
finite-temperature numerical renormalization group 
calculations~\cite{bulla} (restricted to the paramagnetic phase which has a
metal-insulator transition).  We examine three cases here: (i) a 
correlated insulator just above the
transition $U=4.24$ (where the response is ``universal''); 
(ii) a metal just below the phase transition
$U=3.54$ (which undergoes a temperature-dependent metal-insulator 
transition at $T\approx 0.011$); and (iii) a correlated metal $U=2.12$.

\begin{figure}
\epsfxsize=3.0in
\centerline{\epsffile{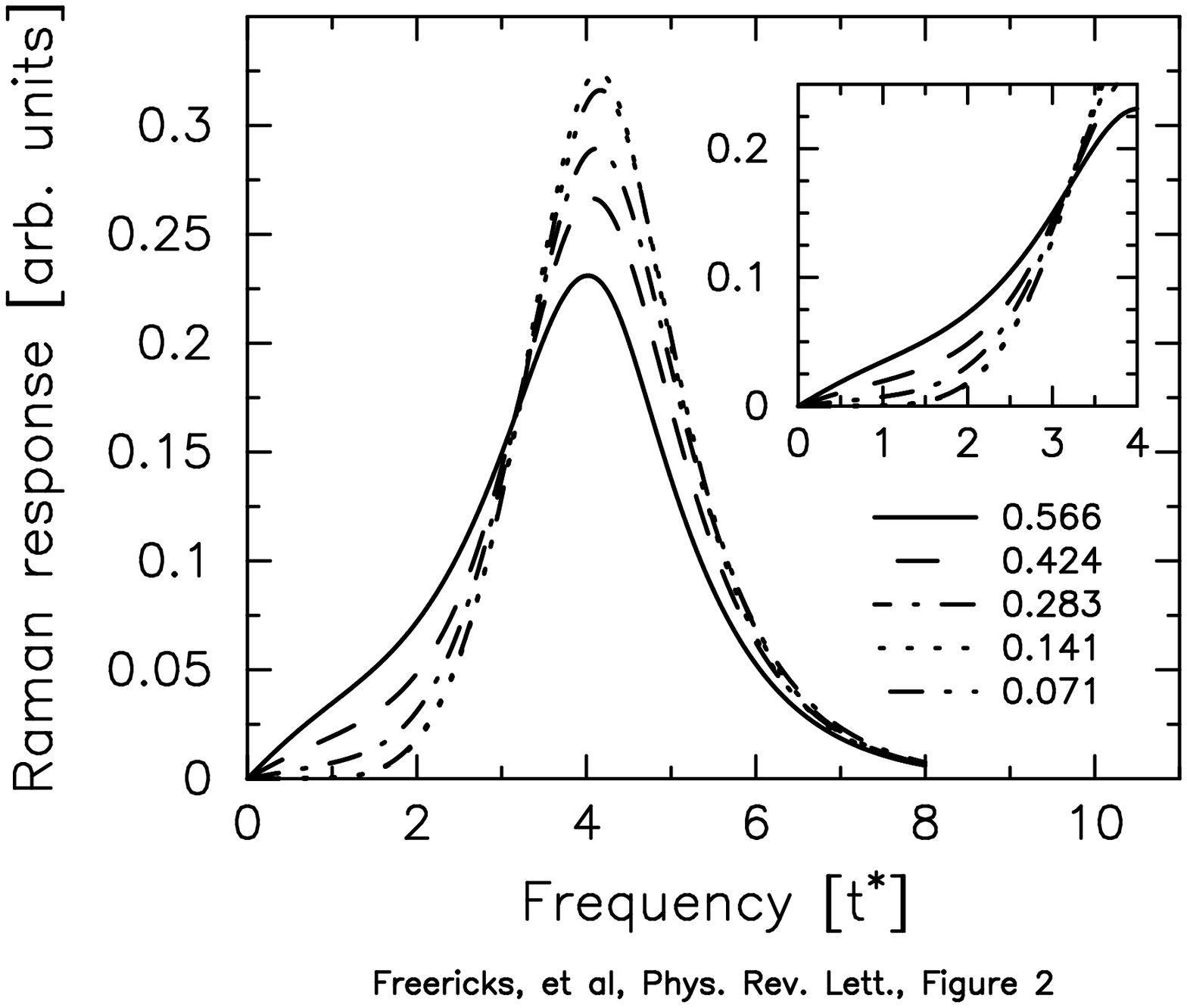}}
\caption{Nonresonant $B_{\rm 1g}$ Raman scattering for a correlated insulator
$U=4.24$ and half filling for a number of temperatures ranging from
$0.57$ to 0.071.  Inset is an enlargement of the low-energy features.
\label{fig: raman_4.24}}
\end{figure}

We show the correlated insulator regime in Fig.~\ref{fig: raman_4.24},
where the Raman response is model-independent.  We see behavior identical to that
seen in the Falicov-Kimball model (the only modification here is the
consequence of a temperature dependence in the interacting
density of states, which fills in the gap at temperatures above about 0.1).
We see the development of low-energy spectral
weight at the expense of the higher-energy charge-transfer peak as $T$ increases
(although in this case the spectral weight grows over a broader temperature
scale) and the appearance of a single isosbestic point at
a somewhat higher energy than seen in the Falicov-Kimball model solution
($\nu\approx U/1.5$ rather than $\nu\approx U/2$). All of the qualitative
features of the universal behavior are shared in the Hubbard-model solution.
The difference is in the quantitative details, the most apparent one being that
at high temperatures the low-energy spectral response does not have a 
broad peak structure, but rather shows monotonic rising behavior.  This
is more characteristic of experiments in materials like FeSi~\cite{SLC2}
or the cuprates~\cite{irwin,hackl,uiuc} which have a relatively flat low-energy
spectral response.

\begin{figure}
\epsfxsize=3.0in
\centerline{\epsffile{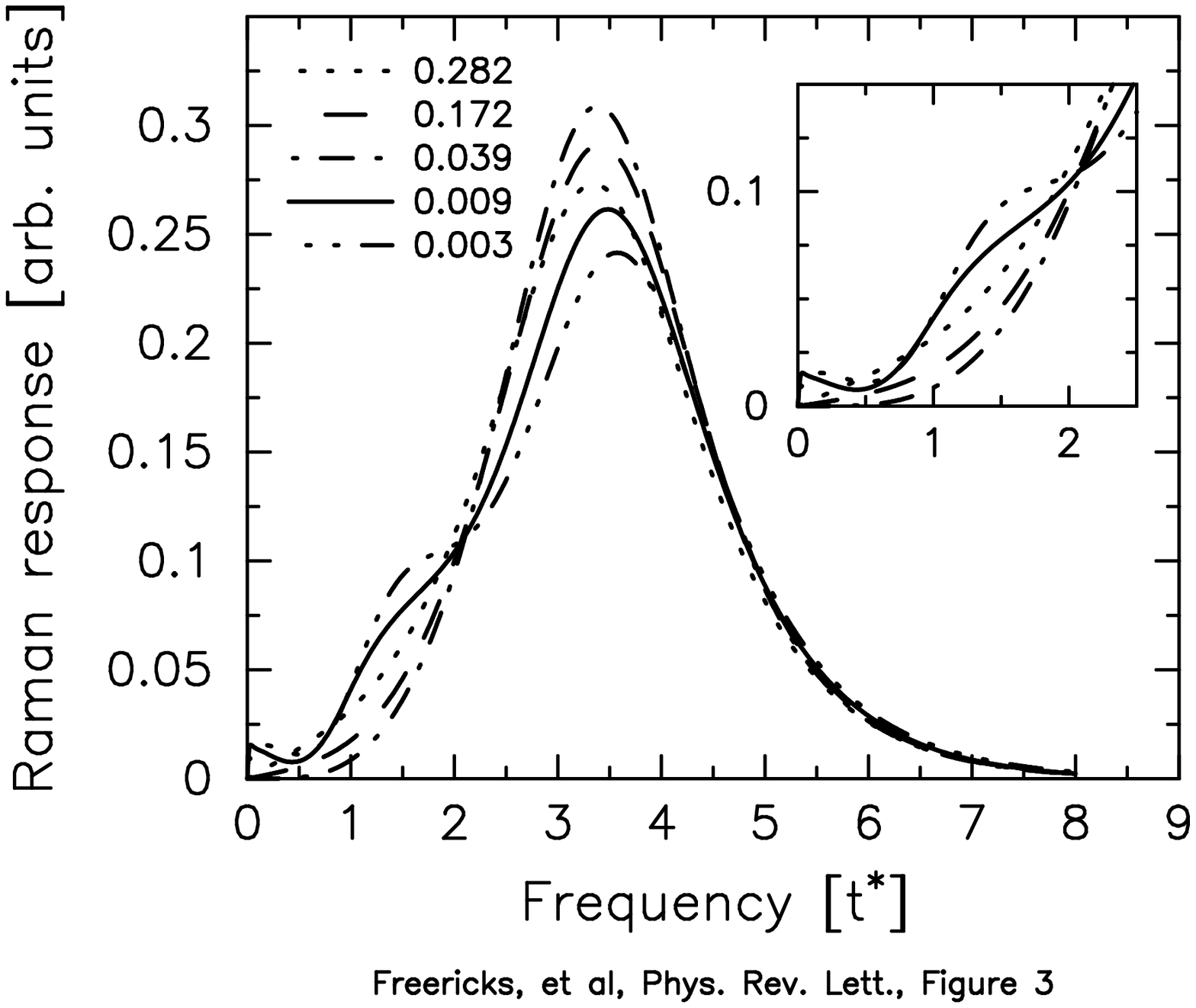}}
\caption{Nonresonant $B_{\rm 1g}$ Raman scattering for system undergoing a
temperature-driven metal-insulator transition $U=3.54$ and half filling
for a number of temperatures ranging from
$0.28$ to 0.003.  Inset is an enlargement of the low-energy features.
\label{fig: raman_3.54}}
\end{figure}

In Fig.~\ref{fig: raman_3.54}, we show results for a system tuned to
lie just on the metallic side of the metal-insulator transition (so
it undergoes a temperature-driven transition at $T\approx 0.011$).
Initially the Raman response acts like an insulator: the charge-transfer
peak sharpens and grows in strength, while the low-energy weight is
reduced and an isosbestic point appears near $\nu=U/1.7$ as $T$ is lowered.  
As the temperature is reduced further, the response becomes quite
anomalous.  Spectral weight shifts back out of the charge-transfer peak
into the low-energy region, developing two low-energy bumps and an isosbestic
point near $\nu=2$.  There is only one experimentally measured
system that we know
of that shares some of these qualitative features: SmB$_6$~\cite{SLC1}.
At low temperatures SmB$_6$ acts like an insulator, just as seen in 
Figure~\ref{fig: raman_3.54}, but then at the lowest temperatures ($T<30~$K), 
a sharp
nondispersive peak appears at about 130~cm$^{-1}$.  The qualitative shape of the
Raman response is different though, because the weight only grows in a narrow
peak, as opposed to the wide frequency range of Figure~\ref{fig: raman_3.54}.
We believe it would be interesting to investigate the Raman response of a
correlated system that undergoes a similar insulator-metal transition as
a function of temperature such as 1\% Chromium doped Vanadium 
Oxide~\cite{vo}.  Such experiments would be feasible if the charge-transfer
peak could be pushed to a low enough energy that it lies within the window
observable by Raman measurements (which seems likely from the optical
conductivity data on insulating Vanadium Oxide alloys doped
close to the metal-insulator transition).

\begin{figure}
\epsfxsize=3.0in
\centerline{\epsffile{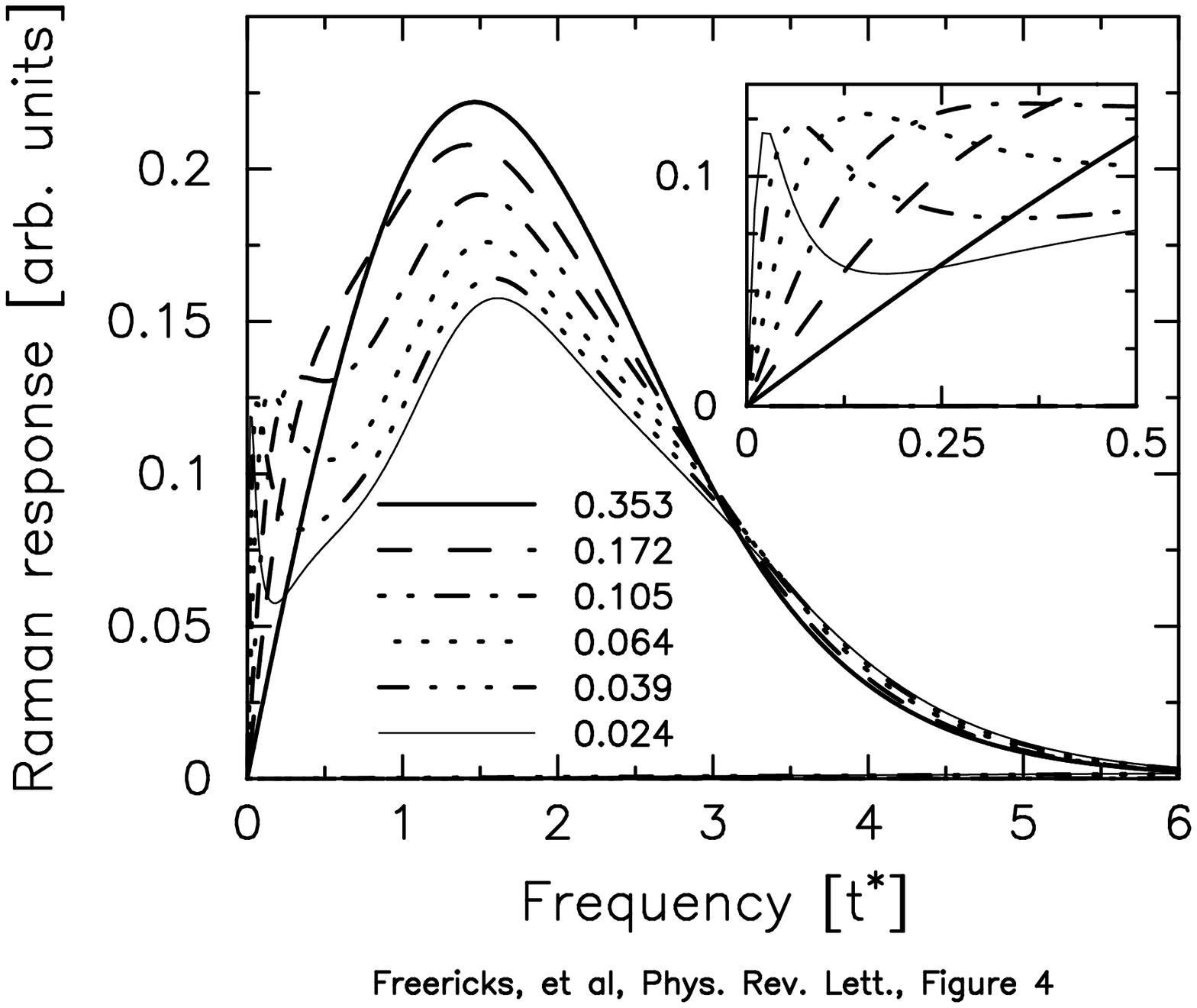}}
\caption{Nonresonant $B_{\rm 1g}$ Raman scattering for a correlated
metal $U=2.12$ at half filling for a number of temperatures ranging from
$0.35$ to 0.024.  Inset is an enlargement of the low-energy features.
\label{fig: raman_2.12}}
\end{figure}

Finally, in Fig.~\ref{fig: raman_2.12}, we show the Raman response for a 
correlated metal at half filling.  At high temperatures, the Raman response has 
a wide charge-transfer peak centered at $\nu\approx U$.  As the temperature is
lowered, we see the development and evolution of a low-energy Fermi-liquid
peak, which sharpens as $T$ is lowered.  This is the classic behavior expected
for a correlated metal---at high temperatures there is a large charge-transfer
peak centered at an energy just somewhat higher than $U$ which
loses spectral weight as the temperature is
lowered and a low-temperature ``metallic'' peak develops at low energy. 
The Fermi peak
has the expected form proportional to $\nu a T/(\nu^2+a^2T^2)$ for a Fermi
liquid.  The width of the peak (determined by $aT$) decreases as the temperature
is lowered and will ultimately vanish at $T=0$.  The weight in the metallic
peak is much smaller than the weight in the charge-transfer
peak.  The continuous evolution of
the low-temperature peak down to zero frequency is not seen in the
Falicov-Kimball model, because the self energy remains finite there in the
limit as $T\rightarrow 0$.  Note that the charge-transfer
peak {\it loses} weight (especially on the low-frequency
side) as $T$ is lowered, but there
is no low-energy isosestic point here (in fact an isosbestic point may be
developing at $\nu\approx 3.3$). This lack of low-energy isosbestic behavior
is quite interesting.  The development of a large-weight Fermi coherence
peak in the interacting density of states tends to destroy the isosbestic
behavior.  

Unfortunately, we are not aware of any experimental data
on a correlated metal that displays this low-temperature development and
evolution of the Fermi-liquid peak.  
Surprisingly, little is known about Raman scattering in a weakly interacting
metal. It is well know that in the absence of inelastic scattering, the low
energy Raman
cross-section vanishes at q=0 due to the lack of phase space available 
to create
particle-hole pairs. Therefore any signal at all must come from 
electron-electron
interactions\cite{weak} or impurities\cite{zawa}. Our results show for the
first time that the consequences of well-defined quasiparticles is the presence
of a low energy peak which grows in intensity and narrows as temperature 
decreases. Such a peak might also be observable
in materials like CeSi$_2$, CeSb$_2$, CeBe$_{13}$ \cite{joyce} or 
YbAl$_3$ \cite{tjeng} which all display the development of a low-temperature
Fermi coherence peak at energies less than 100~meV, but 
may be below the sensitivity of most experiments near the laser line. 

To summarize, 
we have examined the effects of Fermi-liquid behavior in the metallic
phase on the Raman response of a correlated material that is tuned to
lie close to the quantum-critical point of a metal-insulator transition.
In the case of half-filling, where we can tune the system to move right
through the quantum-critical point, we see similar behavior as seen in
the Falicov-Kimball model, but influenced by the appearance of Fermi-liquid
behavior on the metallic side of the transition.  Hence, for a correlated
metal, there is a charge-transfer peak, which has a Fermi-liquid peak separate
from it at low temperatures and frequencies.  The peak sharpens and moves
to lower energy as $T$ is lowered and an isosbestic point may be forming on the
high-energy-side of the charge-transfer peak.  In the insulating regime, we 
recover the universal behavior predicted, which shows a low-temperature 
increase in low frequency spectral weight at the expense of the charge-transfer
peak and the appearance of a single isosbestic point (but with somewhat
stronger temperature dependence).  The quantitative features appear to
be closer to experiment than what was seen in the Falicov-Kimball model.
The region closest
to the quantum-critical point is the most interesting. In the insulating
(high-temperature)
phase it behaves like a correlated insulator, but once it develops a low
temperature Fermi-liquid peak in the density of states, the low frequency
Raman response {\it increases} as $T$ is lowered, and develops a series of
low-energy bumps.  This behavior hasn't yet been seen experimentally,
but it would be interesting to look for in a material that undergoes a
low temperature metal-insulator transition (such as the doped
Vanadium-Oxide system).

\acknowledgments

J.K.F. acknowledges support of the National Science Foundation under grant 
DMR-9973225.  T.P.D. acknowledges support from the National Research and
Engineering Council of Canada.  R.B. acknowledges support by the Deutsche 
Forschungsgemeinschaft, through the Sonderforschungsbereich 484.
We also acknowledge useful discussions with S.L. Cooper, R. Hackl,
J.C. Irwin, M.V. Klein, and S. Shastry. We also thank 
S.L. Cooper, R. Hackl, and J.C. Irwin
for sharing their data with us.

\end{document}